\documentclass[aps,superscriptaddress,reprint,floatfix]{revtex4-1}
\usepackage{bm}
\usepackage{amsmath,amssymb,cases}
\usepackage{graphicx}
\usepackage{xcolor}
\usepackage{soul}
\usepackage{braket}

\begin{document}

\title{Dynamical formation of quantum droplets in a $^{39}$K mixture}
\author{G. Ferioli}
\affiliation{\mbox{LENS and Dipartimento di Fisica e Astronomia, Universit\'a di Firenze, 50019 Sesto Fiorentino, Italy}}
\affiliation{CNR Istituto Nazionale Ottica, 50019 Sesto Fiorentino, Italy}
\author{G. Semeghini}
\affiliation{Department of Physics, Harvard University, Cambridge, MA 02138, USA}
\author{S. Terradas-Brians\'o}
\affiliation{Instituto de Ciencia de Materiales de Arag\'on and Departamento de F\'isica de la Materia Condensada, CSIC-Universidad de Zaragoza, 50009, Zaragoza, Spain}
\author{L. Masi}
\affiliation{\mbox{LENS and Dipartimento di Fisica e Astronomia, Universit\'a di Firenze, 50019 Sesto Fiorentino, Italy}}
\affiliation{CNR Istituto Nazionale Ottica, 50019 Sesto Fiorentino, Italy}
\author{M. Fattori}
\affiliation{\mbox{LENS and Dipartimento di Fisica e Astronomia, Universit\'a di Firenze, 50019 Sesto Fiorentino, Italy}}
\affiliation{CNR Istituto Nazionale Ottica, 50019 Sesto Fiorentino, Italy}
\author{M. Modugno}
\affiliation{\mbox{Departamento de F\'isica Te\'orica e Historia de la Ciencia, Universidad del Pais Vasco UPV/EHU, 48080 Bilbao, Spain}}
\affiliation{IKERBASQUE, Basque Foundation for Science, 48013 Bilbao, Spain}

\date{\today}

\begin{abstract}
We report on the dynamical formation of self-bound quantum droplets in attractive mixtures of $^{39}$K atoms. Considering the experimental observations of Semeghini \textit{et al.}, Phys.  Rev.  Lett. \textbf{120},  235301 (2018), we perform numerical simulations to understand the relevant processes involved in the formation of a metastable droplet from an out-of-equilibrium mixture. We first analyze the so-called self-evaporation mechanism, where the droplet dissipates energy by releasing atoms, and then we consider the effects of losses due to three-body recombinations and to the balancing of populations in the mixture. We discuss the importance of these three mechanisms in the observed droplet dynamics and their implications for future experiments.
\end{abstract}

\maketitle

\section{Introduction}
Self-bound droplets of ultracold atoms have been recently discovered as a new exotic quantum phase \cite{petrov2015,petrov2018liquid,ferrier2018quantum, ferrier2019PhysToday}. Although forming in dilute atomic gases, they display properties which are unique in that context but common to different systems such as classical liquids, helium nanodroplets or atomic nuclei. After their existence in attractive bosonic mixtures was theoretically predicted in \cite{petrov2015}, they have been experimentally observed in dipolar condensates \cite{ferrier2016observation, schmitt2016self, chomaz2016quantum, tanzi2019}, homonuclear mixtures of $^{39}$K \cite{cabrera2018quantum,semeghini2018} and recently in a heteronuclear mixture of $^{41}$K and $^{87}$Rb \cite{derrico2019}.

The first pioneering experiments performed with homonuclear mixtures at ICFO \cite{cabrera2018quantum, cheiney2018bright} and LENS \cite{semeghini2018, ferioli2019collisions} have been able to demonstrate the existence of self-bound droplets in these systems and to provide a first characterization of their peculiar features. While these works have been mainly devoted to study the droplets equilibrium properties, the experiment reported in Ref. \cite{semeghini2018} has also shown that during the formation of the droplet an interesting and complex dynamics takes place. The non-adiabatic preparation of the mixture leads to an initial compression of the atomic cloud and to following oscillations of its size, while the presence of strong three-body losses (3BL) continuously drives the system out of equilibrium. During this evolution, the droplet eventually reaches a metastable state, which was the main focus of the investigation performed in \cite{semeghini2018}. In this work we concentrate instead on the dynamical evolution observed during the droplet formation, trying to understand the different mechanisms involved and their relative importance. While three-body recombinations are a well-known phenomenon in the field of ultracold atoms, the other two mechanisms playing a role in this evolution are specific of quantum droplets and thus require further attention. The first is related with the need to adjust the populations in the components of the mixture to balance its interaction energy. The second, the so-called \textit{self-evaporation}, is a more peculiar dissipation mechanism predicted in Ref. \cite{petrov2015}. Calculating the excitation spectrum of the droplet, Petrov noticed that, in some specific conditions, the droplet cannot host any discrete excitation, since all the excited states are higher in energy than the particle emission threshold. This suggested the idea that the droplet could be able to dissipate any excess of energy by expelling atoms, from which the term \textit{self-evaporation}.

In this paper, we consider the specific experimental case of Ref. \cite{semeghini2018} and use numerical simulations to understand how these different mechanisms come into play in the evolution of the droplet. In Sec.~\ref{sec:selfbound} we summarize the conditions for the existence of self-bound states in a $^{39}$K atomic cloud at zero temperature. In Sec.~\ref{sec:selfevap} we analyze the phenomenon of self-evaporation, considering the ideal case of a mixture without 3BL. We first study the linear regime, where the droplet is prepared with a small initial excitation, compatible with the assumptions of Ref. \cite{petrov2015}, and then we investigate how the concept of self-evaporation extends to the more realistic case where the mixture is prepared far from equilibrium. In Sec.~\ref{sec:dynamics} we introduce 3BL, thus fully recovering the experimental conditions of Ref. \cite{semeghini2018}, and we analyze the dynamical evolution of the droplet, identifying which mechanisms play a key role in the droplet dynamics. Finally, conclusions are drawn in Sec.~\ref{sec:conclusions}.

\section{Self-bound droplets in $^{39}$K}
\label{sec:selfbound}

The key ingredient for the formation of self-bound atomic clouds is the competition between attractive and repulsive forces which, scaling differently with the atomic density, generate a binding potential \cite{petrov2015}. In a mixture of $^{39}$K atoms in the hyperfine states $|1,0\rangle \equiv|1\rangle$ and $|1,-1\rangle \equiv |2\rangle$, this situation occurs in a specific range of magnetic fields $B$, where the intraspecies scattering lengths $a_{11}$ and $a_{22}$ are positive, while the interspecies $a_{12}$ is negative. For $B<56.85$ G, the quantity $\delta a \equiv-|a_{12}|+\sqrt{a_{11}a_{22}}$ becomes negative, so that the global mean-field (MF) interaction is attractive. In this regime, a repulsive effect is provided by quantum fluctuations, corresponding to the so-called Lee-Huang-Yang (LHY) energy term \cite{lee1957}, that stabilizes the system against collapse and gives rise to a self-bound atomic droplet. As derived in Ref. \cite{petrov2015}, the competition between the LHY energy and the attractive MF term locks the ratio between the equilibrium densities in the two species to
\begin{equation}
n_1^{(0)}/n_2^{(0)}=\sqrt{a_{22}/a_{11}},
\label{eq:ratio}
\end{equation}
with
\begin{equation}
n_{i}^{(0)} = \frac{25\pi}{1024}\frac{1}{\sqrt{a_{ii}}}\frac{\delta a^{2}}{a_{11}a_{22}(\sqrt{a_{11}}+\sqrt{a_{22}})^{5}}.
\label{eq:density}
\end{equation}
The experiments reported in Refs. \cite{cabrera2018quantum,semeghini2018} have verified the existence of these self-bound atomic clouds in the predicted interaction regime at the nominal population ratio of Eq.~(\ref{eq:ratio}).

\section{Self-evaporation}
\label{sec:selfevap}
As discussed in Ref. \cite{petrov2015}, for small atom numbers (see later on for a more precise definition) the droplet has no collective modes with energy lower than the particle emission threshold. Therefore, no bound collective excitation can be sustained in that regime, so that any perturbation of the equilibrium state would result in a release of particles or in a break-up of the droplet into smaller pieces. In this sense, a quantum droplet represents a self-evaporating object. In this section we use numerical simulations to characterize this phenomenon, looking at the dynamical evolution of an excited droplet. We first consider the linear regime, where the system is prepared with a small initial excitation, and then we extend our analysis to a regime closer to the experimental conditions, where the droplet is prepared in a highly excited state.

\subsection{Linear regime}
\label{sec:linear}

For small-amplitude excitations, we can describe the system with a single wave function $\phi(\textbf{r},t)$, neglecting any possible relative motion of the mixture components. We use the same formalism of \cite{petrov2015}, which we briefly summarize here.
We introduce the rescaled spatial coordinate $\bm{\rho}=\bm{r}/\xi$, with
\begin{align}
\xi&\equiv
\left[
\frac{384}{25\pi^{2}}\frac{a_{11}a_{22}(\sqrt{a_{11}}+\sqrt{a_{22}})^{6}}{|\delta a|^{3}}
\right]^{1/2},
\end{align}
and the rescaled time $\tau\equiv \hbar t/m \xi^2$.
The droplet wave function $\phi(\bm{\rho},\tau)$ evolves according to the time-dependent Gross-Pitaevskii equation
\begin{equation}
i\partial_{\tau}\phi=\left[-\frac12\nabla_{\rho}^{2} - 3|\phi|^{2} +\frac52|\phi|^{3}\right]\phi,
\label{eq:droplet}
\end{equation}
where $\int|\phi|^{2}d^{3}\rho=\widetilde{N}$ defines the chemical potential $\tilde{\mu}$ and $\widetilde{N}$ is related to the number of atoms in the two atomic species $N_i$ by
\begin{equation}
N_{i}=n_{i}^{(0)}\xi^{3}\widetilde{N},\quad i=1,2.
\label{eq:ntilde}
\end{equation}
It was predicted in Ref. \cite{petrov2015} and confirmed by the experiments in Refs. \cite{cabrera2018quantum,semeghini2018} that stable droplets exist only for $\widetilde{N}>\widetilde{N}_c\approx18.65$. The regime of self-evaporation corresponds to  $20.1<\widetilde{N}<94.2$ \cite{petrov2015}.

In the following, we will restrict our analysis to the case of a spherically symmetric system, which reduces Eq.~(\ref{eq:droplet}) to an effectively one-dimensional equation which depends on the radial coordinate $\rho$ only. This assumption will be maintained throughout this work, for easiness of calculations and conceptual clarity. Note that in this approximation the only possible excitation is that with angular momentum $\ell=0$, i.e., the monopole mode. Higher angular momentum modes are not accounted for, so that the system is self-evaporating up to $\widetilde{N}\simeq934$, where the monopole mode reenters into the spectrum (see Ref. \cite{petrov2015}).

To study the dynamics of an excited droplet, we prepare the system slightly out of equilibrium, by solving the stationary version of Eq.~(\ref{eq:droplet}) with a finite tolerance \footnote{The stationary solutions of Eqs. (\ref{eq:droplet}) and (\ref{eq:gs}) are obtained by means of imaginary time evolution, using a split-step algorithm which includes a Crank-Nicholson propagator for the radial coordinate. A similar approach is employed for computing the real time evolution. See e.g., \cite{Press1996}.}, so that the initial wave function $\phi_0^{(i)}(\rho)$ corresponds to an excited state with a  slightly larger energy than the actual ground state $\phi_0(\rho)$ \footnote{The ground state of the system is computed by propagating a trial wave function in imaginary time, until the variation of the chemical potential in the unit step is below a given \textit{tolerance}. An excited state can be then obtained simply by increasing the value of that tolerance, therefore stopping beforehand the imaginary time evolution.}.
We then calculate the evolution of the system by solving Eq.~(\ref{eq:droplet}) with the above initial condition. To distinguish the droplet from unbound expanding components that may form during the evolution, we define a \textit{droplet volume} as that contained within a certain bulk radius $R_{d}(\tau)$, defined as the position of the minimum of $\bar{n}(\rho,\tau)\equiv \rho^2 n(\rho,\tau)$ (see Appendix \ref{sec:appendix} for details) \footnote{We remark that the numerical simulations of the GP equation are performed on a computational box that is at least one order of magnitude larger than $R_{d}$.}.

\begin{figure}[tb]
\centerline{\includegraphics[width=0.9\columnwidth]{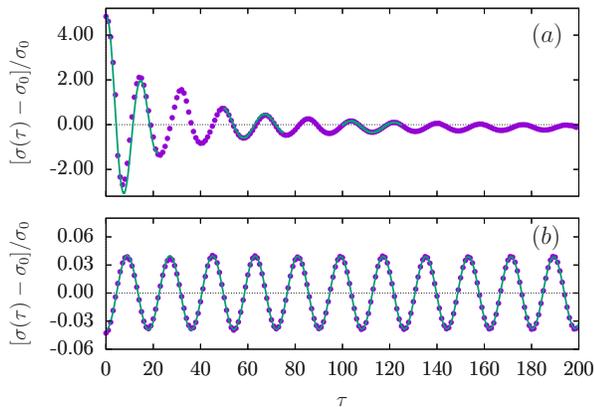}}
\caption{Evolution of the droplet size $\sigma(\tau)$, plotted in terms of its relative variation with respect to the equilibrium value $\sigma_{0}$, for small initial excitations, for $(\widetilde{N} -\tilde{N_c})^{1/4} \simeq 3.5$ (a) and $(\widetilde{N} -\tilde{N_c})^{1/4}\simeq 7$ (b). Purple dots correspond to the values extracted from the numerical simulation, while the green solid lines are the result of damped sinusoidal fits on separate time intervals (a), and of a pure sinusoidal fit (b).
}
\label{fig1}
\end{figure}

The evolution of the droplet size $\sigma(\tau)$, defined as the rms size of the density distribution within the droplet volume, is shown in Fig.~\ref{fig1}.
We find two different behaviors depending on the value of $\widetilde{N}$. For low values of $\widetilde{N}$ [in panel (a)], we observe a damped oscillation with decreasing frequency $\omega_0(\tau)$. In this regime, where all the droplet excitations lie in the continuum, the droplet cannot host a bound monopole mode. This means that when the cloud starts oscillating, part of the atoms move away from it, thus reducing the droplet energy by $\Delta E=-\mu \Delta N +E_{kin}$, where $\Delta N$ is the number of atoms leaving the droplet and $E_{kin}$ is their average kinetic energy. For large $\widetilde{N}$ instead, when the monopole has reentered into the spectrum [panel (b)], the size $\sigma$ performs a sinusoidal oscillation with a well defined frequency $\omega_0$.

\begin{figure}[t]
\centerline{\includegraphics[width=0.9\columnwidth]{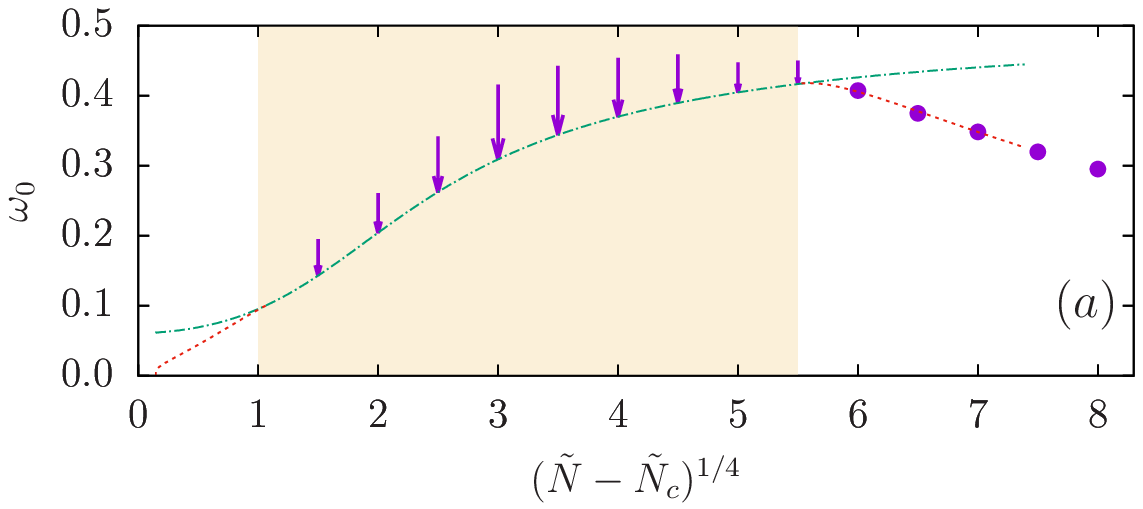}}
\vspace{0.2cm}
\centerline{\includegraphics[width=0.9\columnwidth]{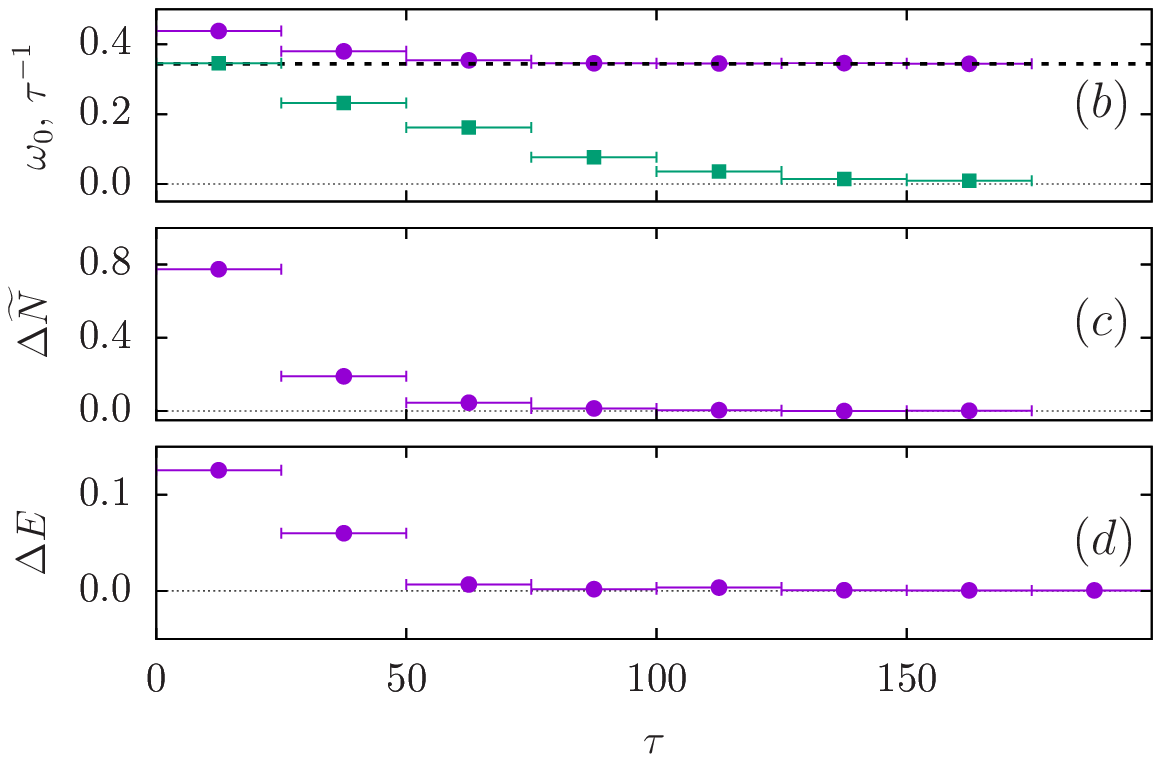}}
\caption{
(a) Plot of the monopole frequency $\omega_0$ (obtained from the numerical simulation) as a function of $(\widetilde{N}-\widetilde{N}_c)^{1/4}$. In the self-evaporation region (shaded area) the range of variation of $\omega_0$ is shown by (purple) arrows. Out of the self-evaporation regime, where the droplet size performs a sinusoidal oscillation as in Fig.~\ref{fig1}b, the value of $\omega_0$ is represented by solid circles. The lines correspond to the theoretical predictions of Ref. \cite{petrov2015} for $-\tilde{\mu}(\widetilde{N})$ (green dashed line) and $\tilde{\omega}_0(\widetilde{N})$ (red dotted line). (b) Values of $\omega_0$ (purple circles) and $1/\tilde{\tau}$ (green squares) fitted on consecutive time intervals of length $\Delta \tau=25$ (indicated as error bars), extracted from the damped oscillation of Fig.~\ref{fig1}a. The black dashed line corresponds to $-\tilde{\mu}$. (c,d) Variation of the parameter $\widetilde{N}$ and of the droplet energy $E$ in each time interval (see text), respectively.
}
\label{fig2}
\end{figure}

To characterize the latter regime, we fit $\sigma(\tau)$ for $\widetilde{N}>940$ with $A \cos(\omega_0 \tau)$ to extract the monopole frequency. For lower $\widetilde{N}$ instead, we fit the oscillation with separate damped cosine functions of the form $A \cos(\omega_0 \tau) \exp{(-\tau/\tilde{\tau})}$ on different intervals of length $\Delta \tau=25$, and we extract the corresponding frequency $\omega_{0}$ and damping rate $\tilde{\tau}$. In Fig.~\ref{fig2}a we report the fitted values of $\omega_0$ for different values of $\widetilde{N}$ and we compare them with the theoretical predictions of Ref. \cite{petrov2015} for $\omega_0 (\widetilde{N})$ and for the emission threshold $-\tilde{\mu}$. For $\widetilde{N}>940$ we find a very good agreement between our numerical results and the prediction for $\omega_0$. In the self-evaporation regime we see that the oscillation frequency decreases in time until it reaches its lower value, set by $-\tilde{\mu}$. In Fig.~\ref{fig2}b we plot the extracted values of $\omega_0(\tau)$ and $\tilde{\tau}(\tau)$ from the fits of Fig.~\ref{fig1}a. In Fig.~\ref{fig2}(c,d) we report the variation of the parameter $\widetilde{N}$ and of the droplet energy $E$ in the corresponding time intervals, namely $\Delta\widetilde{N}(\tau_i)\equiv\widetilde{N}(\tau_{i}-\Delta\tau/2)-\widetilde{N}(\tau_{i}+\Delta\tau/2)$ and $\Delta{E}(\tau_i)\equiv{E}(\tau_{i}-\Delta\tau/2)-{E}(\tau_{i}+\Delta\tau/2)$. We observe that the cloud initially oscillates with large frequencies and fast damping rates, associated with a significant release of atoms from the droplet into an unbound cloud. As the droplet energy decreases due to atom losses, the oscillation slows down and $\omega_0$ eventually saturates at $-\tilde{\mu}$. Close to this threshold the atoms that leave the droplet carry away $\simeq -\tilde{\mu}$ and thus have negligible kinetic energy. The damping of these small final oscillations is then extremely slow and the droplet reaches the stationary ground state only asymptotically. However, these residual excitations are extremely small and one can effectively consider that the droplet has reached its equilibrium configuration well before the asymptotic regime is reached.

\subsection{Non-linear regime}
\label{sec:non-linear}

In the experiment of Ref. \cite{semeghini2018}, the mixture is prepared far from the equilibrium profile $\phi_0$ of a droplet. Then, to understand the dynamics which takes place during the droplet formation we extend the previous analysis beyond the linear case.
Since we cannot exclude \textit{a priori} a different behavior of the two components of the mixture, here we replace Eq.~(\ref{eq:droplet}) with a system of two coupled generalized Gross-Pitaevskii (GP) equations \cite{dalfovo1999} including LHY corrections \cite{semeghini2018,derrico2019} (we keep the assumption of spherical symmetry to simplify the discussion)
\begin{numcases}{}
i\hbar\partial_{t}\psi_{1}= \left[-\frac{\hbar^2\nabla_{r}^{2}}{2m} +\mu_{1}(n_{1},n_{2}) \right]\psi_{1}
\nonumber
\\
\label{eq:2gpe}
\\
i\hbar\partial_{t}\psi_{2}= \left[-\frac{\hbar^2\nabla_{r}^{2}}{2m} +\mu_{2}(n_{1},n_{2}) \right]\psi_{2},
\nonumber
\end{numcases}
where $\psi_{i}(r)$ are the wave functions of the two species and
the corresponding chemical potentials are given by
\begin{equation}
\mu_{i}=g_{ii}n_{i} + g_{12}n_{j} + \frac{\delta E_{LHY}}{\delta n_{i}},\quad i\neq j
\end{equation}
with $g_{ij}=4\pi\hbar^{2}a_{ij}/m$. The LHY energy is \cite{petrov2015}
\begin{equation}
E_{LHY}=\frac{8m^{3/2}}{15\pi^{2}\hbar^{3}}\int_{V} \left(g_{11}n_{1}+g_{22}n_{2}\right)^{5/2},
\label{eq:elhy3}
\end{equation}
so that
\begin{equation}
\frac{\delta E_{LHY}}{\delta n_{i}}=\frac{32}{3\sqrt{\pi}}g_{ii}\left(a_{11}n_{1}+a_{22}n_{2}\right)^{3/2}.
\end{equation}
Note that a similar model, obtained by means of a local density approximation, has been used also for the description of dipolar quantum droplets \cite{wachtler2016a, wachtler2016b, baillie2016self}.

In our numerical simulations we consider a preparation of the mixture similar to the one implemented in the experiment of Ref. \cite{semeghini2018}. A single species Bose-Einstein condensate (BEC) of $N$ atoms of $^{39}$K in the hyperfine level $|2\rangle$ is prepared in the ground state of a harmonic trap with trapping frequency $\omega $. At $t=0$, $N_{1}$ atoms are instantaneously transferred to $|1\rangle$, with $N_{2}=N-N_{1}$ atoms remaining in the initial state. The harmonic potential is then turned off to study the evolution of the mixture in free space. Here, we consider a typical set of experimental parameters, with $N=N_{1}+N_{2}=4\times10^{5}$ and scattering lengths $a_{11} \simeq 69.99 a_{0}$, $a_{12}\simeq-53.37 a_{0}$ and $a_{22}\simeq34.11 a_{0}$, which correspond to a Feshbach magnetic field $B=56.45$ G. For this set of parameters, we have $\widetilde{N}=200$, which lies in the self-evaporation regime identified above.

The ground state wave function $\psi_{0}$ of the initial condensate is obtained from the following stationary GP equation \cite{dalfovo1999}
\begin{equation}
\left[-\frac{\hbar^2}{2m}\nabla_{r}^{2}+ \frac12m\omega^{2}r^{2} +g_{22}N|\psi_{0}|^{2} \right]\psi_{0} = \mu\psi_{0}
\label{eq:gs}
\end{equation}
with $\int |\psi_{0}|^{2}d^{3}r\equiv 1$. Considering the preparation sequence described above, we simulate the instantaneous transfer of atoms in $|1\rangle$ by assuming that the (normalized) wave functions of the two components at $t=0$ are equal to $\psi_0$, with the corresponding densities being $n_i=N_i|\psi_0|^2$. Here we keep the ratio $N_1/N_2$ fixed to the nominal ratio in Eq.~(\ref{eq:ratio}). The following evolution is then obtained by solving Eq.~(\ref{eq:2gpe}).

One can easily guess that the droplet is minimally excited when the trapping frequency $\omega$ is such that the initial atomic distribution is as close as possible to that of the droplet ground state for that value of $\widetilde{N}$, namely $|\psi_{0}(r)|^{2}\approx\xi^{-3}|\phi_{0}(r/\xi)|^{2}$.
To quantify the difference between the initial profile of the condensate and that of the droplet, one can consider, e.g., the relative deviation between the corresponding energies, $\Delta[\widetilde{N},\omega]\equiv{\left|E(\psi_{0})-E(\phi_{0})\right|}/{E(\phi_{0})}$. For the present case, we find that $\Delta[\widetilde{N}=200,\omega]$ is minimized for a trapping frequency $\omega/2\pi\simeq600$ Hz.

\begin{figure}[t]
\centering
\centerline{\includegraphics[width=0.9\columnwidth]{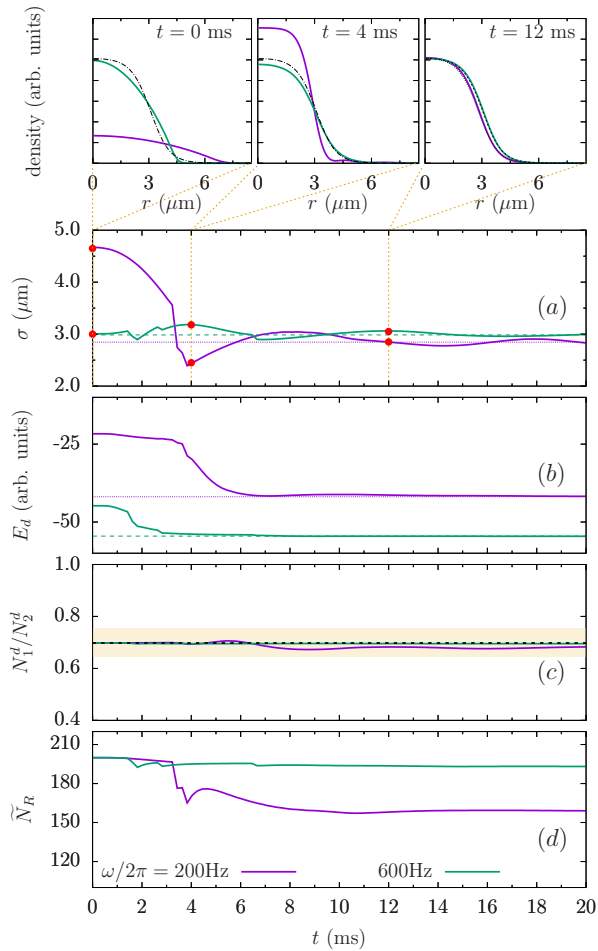}}
\caption{Evolution of the droplet width $\sigma(t)$ (a), its energy $E_{d}(t)$ (b), the ratio $N_{1}^{d}(t)/N_{2}^{d}(t)$ (c), and the running value of $\widetilde{N}_{R}(t)$ (d, see text), for $\omega/2\pi=200$ Hz (purple) and $\omega/2\pi=600$ Hz (green). The insets in the top row show the total density of the binary mixture, $n(r,t)=\sum_{i=1}^{2}N_{i}|\psi_{i}(r,t)|^{2}$, at different evolution times, corresponding to the red circles in (a). As a reference, the profiles of a droplet with $\widetilde{N}=200$ [dot-dashed line, $t=0,4$ ms], $\widetilde{N}=193$ and $\widetilde{N}=159$ [dotted and dashed line respectively, for $t=12$ ms], are also shown in the insets.
In (a,b) the dashed (dotted) horizontal lines represent the equilibrium values of the rms size [in (a)] and the energy for a droplet with $\widetilde{N}=193$ ($159$) [in (b)]. The horizontal line in (c) represent the equilibrium value $N_1/N_2=\sqrt{a_{22}/a_{11}}\simeq0.698$, and the dashed area the corresponding tolerance (see text).}
\label{fig3}
\end{figure}

We then study the dynamics of the system as a function of the distance of the initial state with respect to $\phi_{0}$, by varying the trap frequency $\omega$.
Similarly to the case discussed in the previous section we observe that, after the mixture is formed and released into free space, the atomic clouds undergo damped oscillations.
For $\omega /2 \pi=600$ Hz the density of the binary mixture smoothly adapts to a droplet profile with $\widetilde{N}=193$ in a few milliseconds after the release from the trap (see top insets in Fig.~\ref{fig3}). In this case, the initial excitation energy is very small and indeed we observe a limited dynamics. Remarkably, a droplet is quickly formed even if the trapping frequency differs significantly from the optimal value, as in the case of $\omega/2\pi=200$ Hz (here with $\widetilde{N}=159$), which highlights the efficiency of the self-evaporation mechanism to dissipate the initial excitation energy.

To characterize the relaxation process towards a stationary droplet, here we indicate with $N_{i}^{d}(t)$ ($i=1,2$) the number of atoms remaining within the droplet volume and we define a \textit{running} value of $\widetilde{N}$ from Eq.~(\ref{eq:ntilde}) as
\begin{align}
\widetilde{N}_{R}(t)
&\equiv\frac{5\pi^{2}}{3\sqrt{6}}
\frac{|\delta a|^{5/2}}{(\sqrt{a_{11}}+\sqrt{a_{22}})^{5}}
\left[N_{1}^{d}(t)+N_{2}^{d}(t)\right]
\nonumber\\
&\simeq 0.5\times 10^{-3}\sum_{i=1}^{2}N_{i}^{d}(t)
\end{align}
where the second (approximate) equality holds for the current values of the scattering lengths. We indicate with $\sigma(t)$ the droplet size (evaluated as the average rms of the density distributions of the two atomic components defined as in the previous section, see Appendix \ref{sec:appendix}), and with $E_{d}(t)$ the corresponding energy.
The evolution of $\sigma(t)$, $E_{d}(t)$, $N_{1}^{d}(t)/N_{2}^{d}(t)$ and of the running value of $\widetilde{N}_{R}(t)$ are plotted in Fig.~\ref{fig3}, for $\omega/2\pi=200$ Hz and $600$ Hz. We find that the relaxation process occurs via a sudden expulsion of atoms (at about $t=2\div4$ ms, depending on $\omega$), which allows to dissipate most of the initial excitation energy.
From Fig.~\ref{fig3}b we can estimate an average dissipation rate, by evaluating the decrease in energy occurring in the first part of the evolution, until the deviation from the equilibrium droplet energy becomes very small. We find that it varies from $0.3$ to $2.5$ MHz/ms, as $\omega /2 \pi$ goes from $600$ Hz to $200$ Hz.
Figure \ref{fig3}c shows that the ratio between the atom numbers in the two components always remains very close to the nominal equilibrium value.
We recall that a droplet can sustain an excess of particles in one of the two components
$\delta N_{i}/N_{i}$ up to a critical value $\sim |\delta a|/a_{ii}$, beyond which the particles in excess are expelled \cite{petrov2015}. Here the deviations of $N_{1}^{d}(t)/N_{2}^{d}(t)$ are always within this tolerance (shaded area in Fig.~\ref{fig3}c).

We can conclude that -- in the absence of 3BL -- the relaxation dynamics of a binary mixture produced by means of the non-adiabatic experimental protocol of \cite{semeghini2018} is dominated by the self-evaporation mechanism, consisting in the dissipation of the initial excitation via the release of particles wave packets emitted from the droplet (see also Fig.~\ref{figA1} in Appendix \ref{sec:appendix}).

\section{Dynamics of the droplet formation in the presence of 3BL}
\label{sec:dynamics}

Having now a clear idea of how self-evaporation works, we can discuss the dynamical formation of the droplet in the realistic experimental conditions of Ref. \cite{semeghini2018}, where a significant role is played by 3BL. To do this, we use the same model as in Sec.~\ref{sec:non-linear}, adding to each equation in (\ref{eq:2gpe}) a non-unitary term
\begin{equation}
\left[-i \hbar \frac{K_{iii}}{2} n_{i}^{2}\right]\psi_{i}
\end{equation}
where $K_{iii}$ are the intra-species 3BL rates \footnote{We neglect here the effect of inter-species 3BL, following the experimental observations discussed in Refs. \cite{semeghini2018, cabrera2018quantum}.}.
Their values are ${K_{111}'}/3!=9\times 10^{-40}$ m$^6$/s, ${K_{222}'}/3!=1 \times 10^{-41}$ m$^6$/s, where the primed values correspond to the loss rates of thermal atoms and the factor $1/3!$ accounts for the Bose statistics of condensed atoms \cite{altin2011}.
Following Ref. \cite{semeghini2018}, we assume that the two hyperfine levels are equally populated initially, $N_1= N_2 = 2 \times 10^5$, and we fix the trapping frequency in Eq.~(\ref{eq:gs}) to a value of the same order of the geometric average of the experimental frequencies, namely $\omega=2 \pi \times 200$ Hz.

In the present case, owing to the complex dynamics that originates from the presence of 3BL,
especially in the first part of the evolution, we use a different strategy to distinguish the droplet from unbound expanding atoms, similar to the protocol implemented in \cite{semeghini2018}. We integrate the 3D density profiles along one direction and we fit the column-density with a double gaussian function
$f(r)=A \exp[-r^{2}/(2\sigma^{2})]+B \exp[-r^{2}/(2\sigma_{exp}^{2})]$,
where the first gaussian corresponds to the central droplet and the second one to the unbound expanding cloud.

\begin{figure}[t]
\centering
\centerline{\includegraphics[width=0.9\columnwidth]{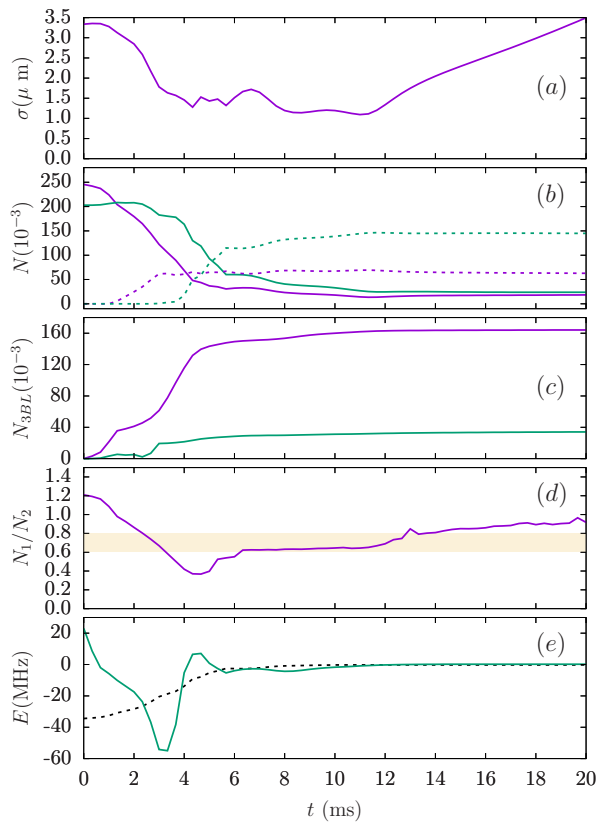}}
\caption{Numerical results for the dynamical formation of the droplet in presence of strong 3BL. (a) Evolution of the droplet size $\sigma(t)$, (b) of the number of atoms in the droplet (solid) and in the unbound cloud (dashed), (c) of the number of atoms lost due to 3BL, (d) of the population ratio in the droplet $N_1/N_2$, and (e) of the droplet energy. In (b,c) we plot the atom numbers separately for the two atomic species $|1\rangle$ (purple) and $|2\rangle$ (green). In (d) we compare the population ratio with the nominal range where the interactions energies in the droplet are properly compensated, $N_1/N_2\simeq\sqrt{a_{22}/a_{11}}$ (shaded area) \cite{petrov2015}. In (e) the droplet energy is shown along with the ground state energy for the corresponding atom number $E_{eq}(N(t))$ (black dashed line).}
\label{fig4}
\end{figure}
In Fig.~\ref{fig4} we show the evolution of the droplet size $\sigma(t)$, of the atom number in the droplet and in the expanding cloud surrounding it, of the droplet population ratio $N_1/N_2$, and finally of the droplet energy. In Fig.~\ref{fig4}e we compare the measured droplet energy with the energy of the ground state for the same atom number $E_{eq}(N(t))$. In the first stages of the evolution, the dynamics is dominated by losses of atoms from state $|1\rangle$, which bring the ratio $N_1/N_2$ closer to the nominal value of Eq.~(\ref{eq:ratio}) and thus significantly reduce the droplet energy. These losses come both from 3B recombinations and from the release of atoms from the droplet into the unbound component, due to the imbalance in the interaction energies. In this stage, the system dissipates the exceeding energy pretty quickly: We measure an energy loss rate of about $20$ MHz/ms. Notice that this is almost $10$ times larger than the self-evaporation cooling rate measured in the previous section for the same initial $\omega$. When the ratio $N_1/N_2$ reaches its nominal value anyway, the droplet keeps losing atoms in $|1\rangle$ due to 3BL. To compensate for that, the droplet starts releasing atoms in $|2\rangle$. This population dynamics, together with the compression of the atomic cloud, causes the two apparent bumps in Fig.~\ref{fig4}e. At this point we find the system close to its equilibrium configuration, with the central cloud forming a metastable droplet with an atom number significantly smaller than the initial one. Since the system keeps losing atoms, due to 3BL in $|1\rangle$ and to population re-equilibration in $|2\rangle$, we see that around $t=12$ ms the droplet population drops below the critical value $\widetilde{N}_c$ \cite{petrov2015, semeghini2018, cabrera2018quantum}, so that the binding mechanism breaks and the cloud starts expanding.

From this analysis we can conclude that the energy variations driven by 3BL and the corresponding population balancing occur on timescales much shorter than those typical of the self-evaporation mechanism described above, so that it is hard to determine if the latter occurs at all. Even if that were the case, its effect would be negligible with respect to the two leading loss mechanisms.

We have also investigated whether a different choice of the initial parameters,
such as the trapping frequency or the population ratio $N_1/N_2$, would lead to a faster equilibration time and/or to longer droplet lifetimes. We find that the values of $\omega$ and $N_1/N_2$ discussed above, which correspond to the values chosen in the experiment \cite{semeghini2018} by optimizing the droplet lifetime, represent indeed a convenient choice. In fact, we observe that a larger value of the initial trapping frequency $\omega$ leads to stronger 3BL in the first stages of the evolution, thus reducing even further the droplet lifetime. On the other hand, a ratio $N_1/N_2$ closer to its equilibrium value $\sqrt{a_{22}/a_{11}}$ does not help either, because the stronger losses in $|1\rangle$ quickly drive the system away from this equilibrium condition, triggering an earlier release of atoms from state $|2\rangle$.

To study experimentally the effect of self-evaporation, one would need to find a suitable attractive mixture with significantly lower 3BL. This might be the case of the $^{41}$K-$^{87}$Rb mixture of the recent experiment in Ref. \cite{derrico2019}, where  thanks to the different values of the scattering lengths the droplet forms at lower densities, so that 3BL are less effective. Indeed, there the lifetime is expected to be one order of magnitude larger than that of the $^{39}$K mixture considered here, at least. In this case, as discussed in Sec.~\ref{sec:non-linear}, it is preferable to prepare the initial state at the nominal equilibrium ratio $N_1/N_2$ and as close as possible to the droplet density profile, to have short equilibration times and limited overall dynamics (see Fig.~\ref{fig2}). Remarkably, if the droplet is prepared in the self-evaporation regime, any unwanted excitation would be then efficiently dissipated via that mechanism.

\section{Conclusions}
\label{sec:conclusions}

We have characterized, by means of numerical simulations, the dynamics of self-evaporation of an atomic quantum droplet, considering both the case of small and large initial excitations. We have verified that the regime  of its occurrence corresponds to that predicted in Ref. \cite{petrov2015}, and we have discussed how and on which timescales this mechanism allows the dissipation of energy in the droplet. We have then simulated the same preparation of the droplet realized in Ref. \cite{semeghini2018}, including the effect of 3BL. This has allowed to identify the relevant mechanisms involved in the dynamical formation of a self-bound droplet of $^{39}$K. In this case, the evolution of the system is dominated by the presence of 3BL and by a continuous release of atoms to restore the proper population ratio, whereas the self-evaporation mechanism plays -- if any -- only a negligible role. We have finally discussed the optimal strategies for the preparation of droplets close to their ground state, both for the experimental case considered here and for possible future experiments with reduced 3BL.
The results reported here, which provide an accurate
characterization of the dynamical evolution of excited droplets, are both useful for a deeper understanding of the recent experimental observations of quantum droplets and relevant for future experimental studies, especially regarding the effect of self-evaporation.

\begin{acknowledgments}
We acknowledge insightful discussions with D. Petrov and E. Sherman. This work was supported by the Spanish Ministry of Science, Innovation and Universities and the European Regional Development Fund FEDER through Grant No. PGC2018-101355-B-I00 (MCIU/AEI/FEDER, UE), the Basque Government through Grant No. IT986-16, the EC-H2020 Grant QUIC No. 641122, and by the project TAIOL of QuantERA ERA-NET Cofund in Quantum Technologies (Grant Agreement N. 731473) implemented within the European Union's Horizon 2020 Programme.
\end{acknowledgments}

\bibliographystyle{apsrev4-1}

%

\begin{appendix}

\section{Defining the droplet radius}
\label{sec:appendix}

\begin{figure}[b!]
\centerline{\includegraphics[width=0.9\columnwidth]{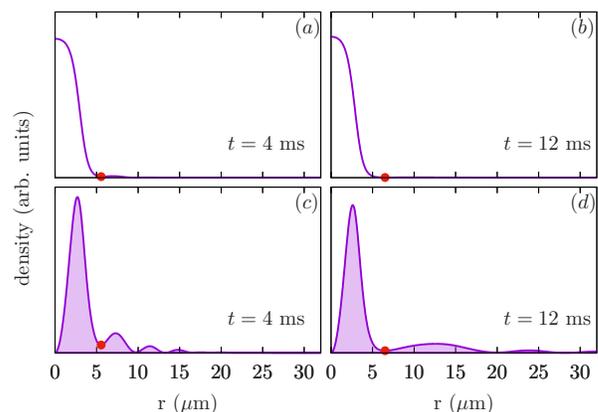}}
\caption{Plot of the total density $n(r,t)=\sum_{i=1}^{2}N_{i}|\psi_{i}(r,t)|^{2}$ (a,c) and of the quantity $\bar{n}(r,t)\equiv r^{2}n(r,t)$, at different evolution times ($t=4,12$ ms -- here for the case $\omega/2\pi=200$ Hz of the simulations discussed in Sec.~\ref{sec:non-linear}). The abscissa of the red dot (see text) represents the value of the droplet radius $R_{d}(t)$.}
\label{figA1}
\end{figure}
The dynamical formation of stationary (or metastable) droplets occurs via a relaxation process during which part of the atoms are expelled from the central volume. Therefore, a criterion for distinguishing between the droplet and the outgoing density waves, is needed. Here we exploit the radial symmetry of the present problem and we consider -- along with the proper density distribution $n(r,t)$ in radial coordinates (Fig.~\ref{figA1}a,b) -- the quantity $\bar{n}(r,t)\equiv r^{2}n(r,t)$ which includes the factor $r^{2}$ corresponding to the Jacobian determinant in spherical coordinates (Fig.~\ref{figA1}c,d). Therefore, a linear integration of $\bar{n}(r,t)$ corresponds to a volume integration of the density $n(r,t)$. This representation also provides a convenient way to visualize the point at which the outgoing density waves ``detach'' from the inner volume, which we identify with the first minimum of $\bar{n}(r,t)$ at the right-hand side of the bulk, marked by a red dot in Fig.~\ref{figA1}c,d. The abscissa of this point is the value that we identify with the droplet radius $R_{d}(t)$. In the numerical simulations of Sec.~\ref{sec:non-linear} for example, the droplet radius is initialized at $R_{d}(0)=10$ $\mu$m and it then evolves -- both continuously or via sudden jumps -- following the position of the leftmost minima (outside of the origin). The latter is typically found between $5$ $\mu$m and $10$  $\mu$m during the whole evolution considered here (up to $20$ ms), corresponding to the fact that the bulk hosts a breathing mode and it does not expand. Contrarily, the density waves which are expelled from the droplet move outward while expanding, as shown in Fig.~\ref{figA1}c,d.

\end{appendix}

\end{document}